\documentclass[conference,10pt]{IEEEtran}

\IEEEoverridecommandlockouts

\usepackage{amsmath,amssymb,amsfonts}
\usepackage{algorithmic}
\usepackage{graphicx}
\usepackage{textcomp}
\usepackage{xcolor}
\usepackage[capitalise]{cleveref}
\usepackage{siunitx}
\usepackage{subfig}

\definecolor{blue}{rgb}{0.0, 0, 0}

\usepackage[backend=biber, sorting=none, bibstyle=ieee,citestyle=numeric-comp]{biblatex}

\small{
    \bibliography{references.bib}
}    
\DefineBibliographyStrings{english}{%
  mathesis = {MSc. Thesis},
}

\begin{document}

\title{Obstacle-Aware Positioning of a Mobile Robotic Platform for 6G Networks
}

\author{
	\IEEEauthorblockN{
	Alexandre Costa, Pedro Duarte, André Coelho, Rui Campos}
	\IEEEauthorblockA{INESC TEC and Faculdade de Engenharia, Universidade do Porto, Portugal\\
    up202005259@edu.fe.up.pt, up202007973@edu.fe.up.pt, andre.f.coelho@inesctec.pt, rui.l.campos@inesctec.pt}
}

\maketitle

\begin{abstract}
The 6G paradigm and the massive usage of interconnected wireless devices introduced the need for flexible wireless networks. A promising approach lies in employing Mobile Robotic Platforms (MRPs) to create communications cells on-demand. The challenge consists in positioning the MRPs to improve the wireless connectivity offered. This is exacerbated in millimeter wave (mmWave), Terahertz (THz), and visible light-based networks, which imply the establishment of short-range, Line of Sight (LoS) wireless links to take advantage of the ultra-high bandwidth channels available.

This paper proposes a solution to enable the obstacle-aware, autonomous positioning of MRPs and provide LoS wireless connectivity to communications devices. It consists of 1) a Vision Module that uses video data gathered by the MRP to determine the location of obstacles, wireless devices and users, and 2) a Control Module, which autonomously positions the MRP based on the information provided by the Vision Module. The proposed solution was validated in simulation and through experimental testing, showing that it is able to position an MRP while ensuring LoS wireless links between a mobile communications cell and wireless devices or users.
\end{abstract}

 
\section{Introduction\label{sec:Introduction}}

Wireless network infrastructures play a crucial role in society, influencing how people communicate, work, and interact with the world. At the same time, the 6G paradigm is emerging, envisioning a massive usage of immersive applications and interconnected devices, including wireless sensors, wireless actuators, wearables, remotely controlled robots, and autonomous vehicles, in different scenarios \cite{Wang2023, Romeo2020}. In order to enable communications in emerging networking scenarios, it is important to ensure that wireless networks are able to accommodate the ever-increasing Quality of Service (QoS) requirements \cite{Vermesan2020}. However, in certain scenarios, establishing resilient and reliable wireless networks is challenging, especially when there is a large number of devices seeking wireless connectivity or when the wireless network infrastructures are destroyed. A reference example is a temporary event, including an emergency management scenario, which occurs during confined periods of time, making the deployment of static wireless network infrastructures not cost-effective.\looseness=-1 

\begin{figure} 
	\begin{center}
		\includegraphics[width=1\columnwidth]{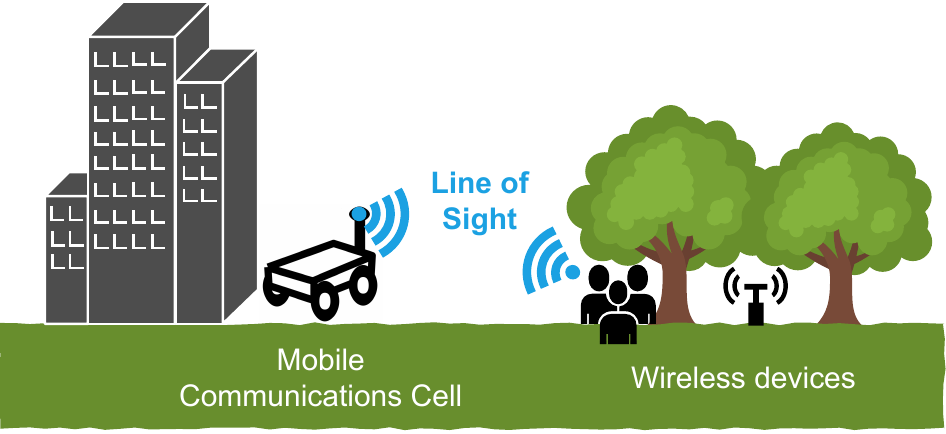}
		\caption{A mobile communications cell providing Line of Sight (LoS) connectivity to wireless devices.}
		\label{fig:cenario}
	\end{center} 
\end{figure}

In order to address the communications challenges in emerging use cases, the integration of Mobile Robotic Platforms (MRPs) into wireless networks is being considered. MRPs, carrying communications nodes on board, enable mobile communications cells capable of establishing, restoring, and reinforcing wireless connectivity dynamically \cite{Maia2022}. However, a significant challenge lies in positioning the MRPs to ensure Line of Sight (LoS) between the mobile communications cells and wireless devices so that the QoS offered is maximized. This is especially relevant when using millimeter wave (mmWave), Terahertz (THz), and visible light-based communications, which enable ultra-high bandwidth channels if short-range, LoS wireless links are ensured. A representative scenario is depicted in \cref{fig:cenario}.\looseness=-1 

The main contribution of this paper is a framework for the obstacle-aware positioning of an MRP to enable LoS communications links in wireless networks, including 1) a Vision Module, able to determine the location of obstacles and devices in the environment surrounding the MRP, using video information collected by the platform; and 2) a Control Module, able to autonomously position the MRP, considering the location of obstacles and wireless devices, determined by the Vision Module. The proposed framework enables the obstacle-aware positioning of an MRP, in order to ensure LoS wireless links between the mobile communications cell and the wireless devices.\looseness=-1 

The rest of this paper is organized as follows. 
\cref{sec:SystemDesign} details the system design and specification, including the novel Vision Module and Control Module proposed. 
\cref{sec:SystemValidation} explains the evaluation carried out, including the obtained results.
\cref{sec:Discussion} discusses the pros and cons of the proposed solution.
Finally, \cref{sec:Conclusions} refers to the main conclusions and future work.\looseness=-1 

\section{System Design and Specification\label{sec:SystemDesign}}

\begin{figure} 
	\begin{center}
		\includegraphics[width=0.4\columnwidth]{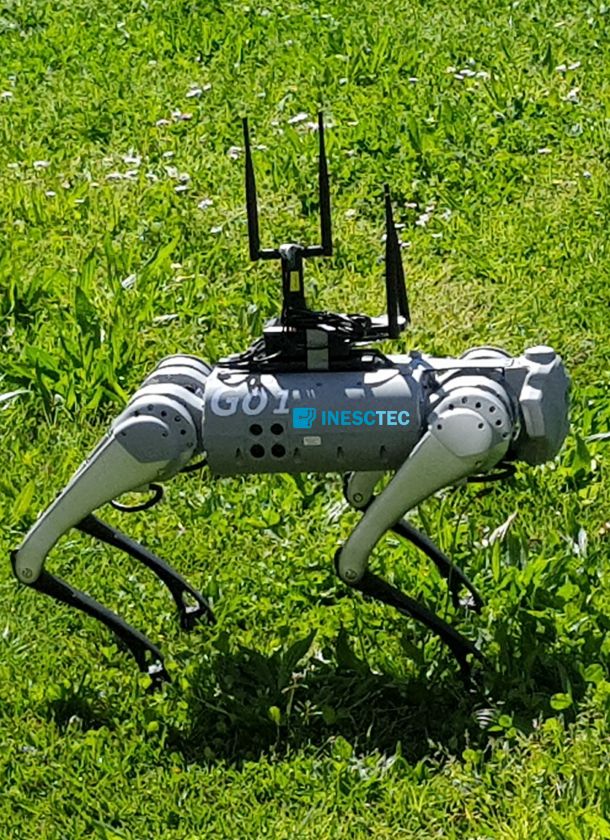}
		\caption{Unitree Go1 mobile robotic platform.}
		\label{fig:GO1}
	\end{center} 
\end{figure}

The MRP used in this research work was the Unitree Go1 \cite{UnitreeGo1}. It is a quadruped robot, shown in \cref{fig:GO1}, especially suitable for research and development of autonomous systems in the fields of human-robot interaction, Simultaneous Localization And Mapping (SLAM), and transportation. The Unitree Go1 platform is targeted for both industry and academia, enabling the development of applications ranging from robotics, wireless communications, to autonomous surveillance. The robot is equipped with multiple sensors, including stereo video cameras, which allow characterizing the surrounding environment. Additionally, it includes three System-on-Module (SOM) devices equipped with a Graphics Processing Unit (GPU) and Central Processing Unit (CPU), which are responsible for processing and transmitting the data captured by the video cameras.\looseness=-1 

\begin{figure} 
	\begin{center}
		\includegraphics[width=0.8\columnwidth]{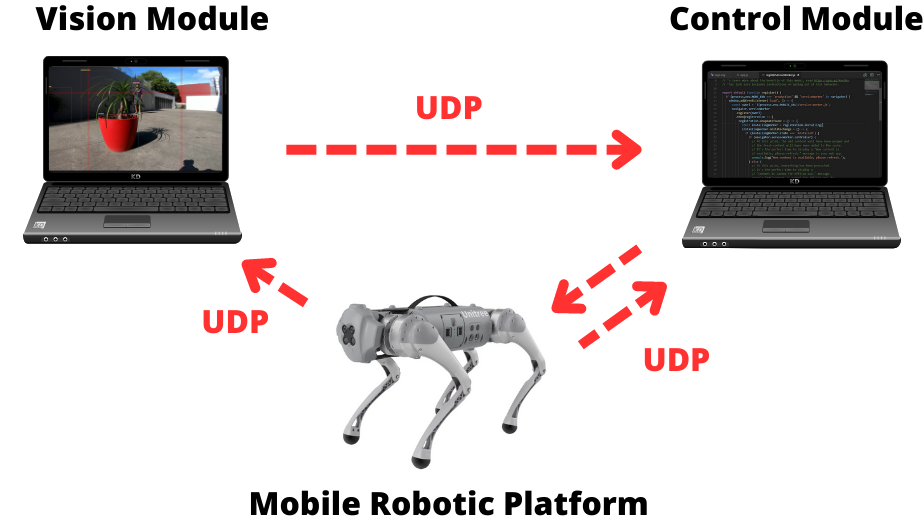}
		\caption{High-level system architecture.}
		\label{fig:testbed_real}
	\end{center} 
\end{figure}

The Unitree Go1 platform is also equipped with a Raspberry Pi, which acts as the main CPU and control board, and three Jetson Nano devices, used to collect the video cameras data. Two laptops were also used to run the Vision Module and the Control Module, respectively. The high-level system architecture is depicted in \cref{fig:testbed_real}. \looseness=-1  

The framework designed for the obstacle-aware positioning of the MRP consists of two modules: 1) a Vision Module that uses the video data to characterize the environment surrounding the MRP; and 2) a Control Module that, based on the information provided by the Vision Module, controls the positioning of the MRP, in order to assure LoS wireless connectivity.\looseness=-1 

\subsection{Vision Module}
\begin{figure}
    \centering
    \includegraphics[width=0.7\columnwidth]{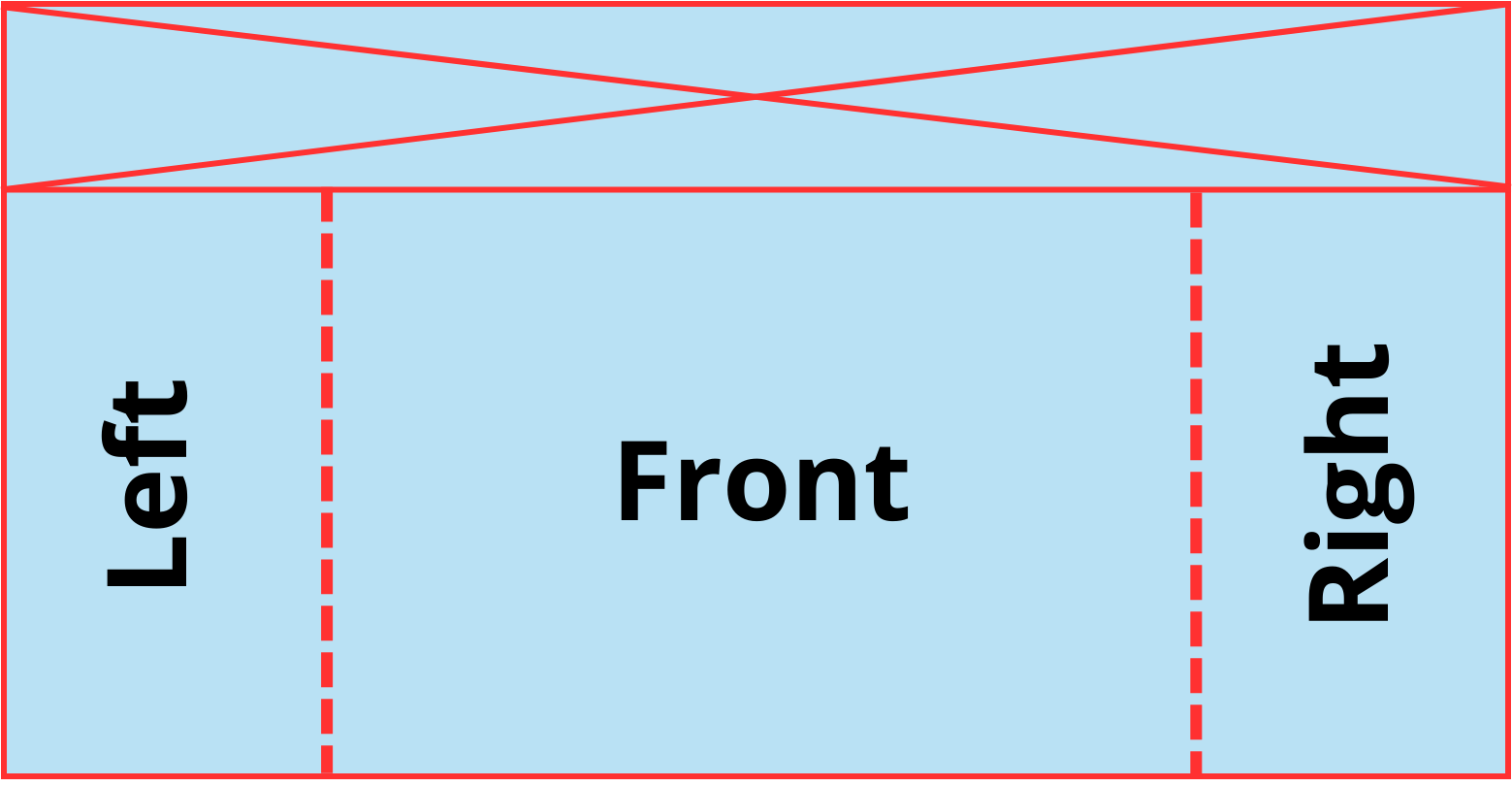}
    \caption{Graphical representation of the detection areas considered by the Vision Module.}
    \label{fig:limites}
\end{figure}

The Vision Module allows us to identify objects in each video frame and classify them as an obstacle or a device. Based on this classification, the position of each object is also determined, considering left, right, and front reference areas, as shown in \cref{fig:limites}. These areas were defined considering the possible directions for the movement of the MRP. For that purpose, the image's height was restricted to 75\% of the total height, while a margin of 25\% was applied on both sides with respect to the width dimension.\looseness=-1 

For classification, the objects are divided into two groups: obstacles and devices. For simplicity, it was assumed that the devices are represented by people, which carry personal wireless devices (e.g., first responders using a body monitoring system). The state of the art YOLO algorithm \cite{yolov7} was used to perform detection and classification of objects in the video frames.\looseness=-1  

\begin{figure}
    \centering
    \includegraphics[width=0.8\columnwidth]{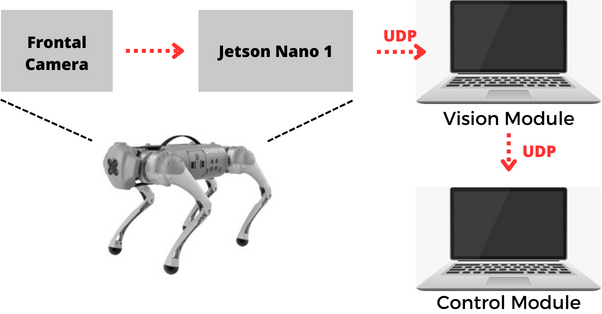}
    \caption{Vision Module's system architecture, including the connection established with the Control Module.}
    \label{fig:vision_module}
\end{figure}

\begin{figure}
    \centering
    \includegraphics[width=0.7\columnwidth]{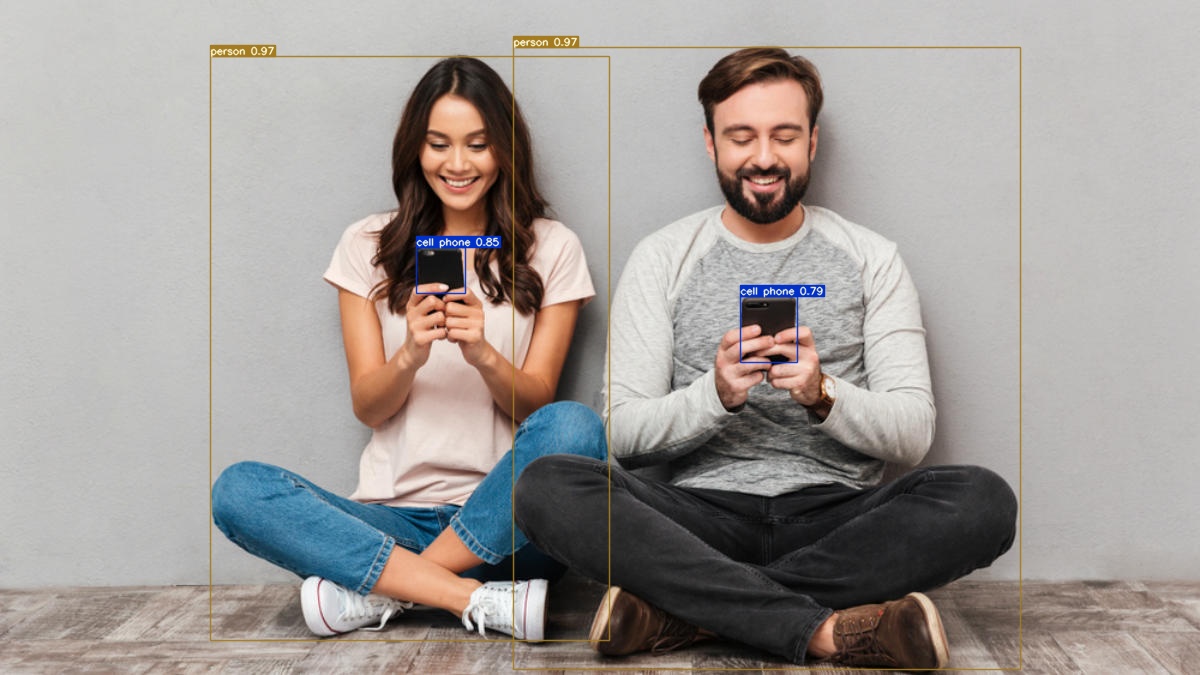}
    \caption{Representative results achieved by the Vision Module, using YOLO \cite{yolov7}.}
    \label{fig:pessoas}
\end{figure}

Considering the information about the distance between the MRP and the objects, estimated by means of ultrasonic sensors, the Vision Module determines if each detected object is close enough to be considered either an obstacle for wireless communications or an wireless device in a suitable position to be served. Finally, the information about the positions of the obstacles and devices is sent to the Control Module using an UDP-based connection, considering the system illustrated in \cref{fig:vision_module}. In order to validate the object detection and classification performed by the Vision Module, extensive testing, considering multiple images, was carried out, as depicted in \cref{fig:pessoas} for a representative example.\looseness=-1 

\subsection{Control Module}
The Control Module considers Cartesian coordinates as input data to move the MRP to a targeted position, assumed to be the location of a device seeking wireless connectivity, while avoiding collisions when obstacles are detected. For that purpose, the Control Module considers the MRP's center as the reference position for movement control. It periodically calculates the orientation angle and distance between the MRP and the targeted position, based on the current position and orientation of the MRP. The Control Module then rotates the MRP to the computed orientation and controls it to move forward in the direction of the targeted position during a predefined time interval, which is a configuration parameter. If an obstacle is detected along the movement trajectory, the MRP stops, and the obstacle avoidance mechanism is activated by the Control Module. The MRP rotates around its center until no obstacles are detected in its field of view and resumes its movement forward. If the movement is successful during the predefined time interval, the MRP recalculates the trajectory to the destination and resumes the movement. Otherwise, the obstacle avoidance function is recursively called. This process is repeated until the MRP reaches the targeted position.\looseness=-1 

\begin{figure}[!ht]
    \centering
        \subfloat[\centering Function $forward$.]{{\includegraphics[width=0.7\columnwidth]{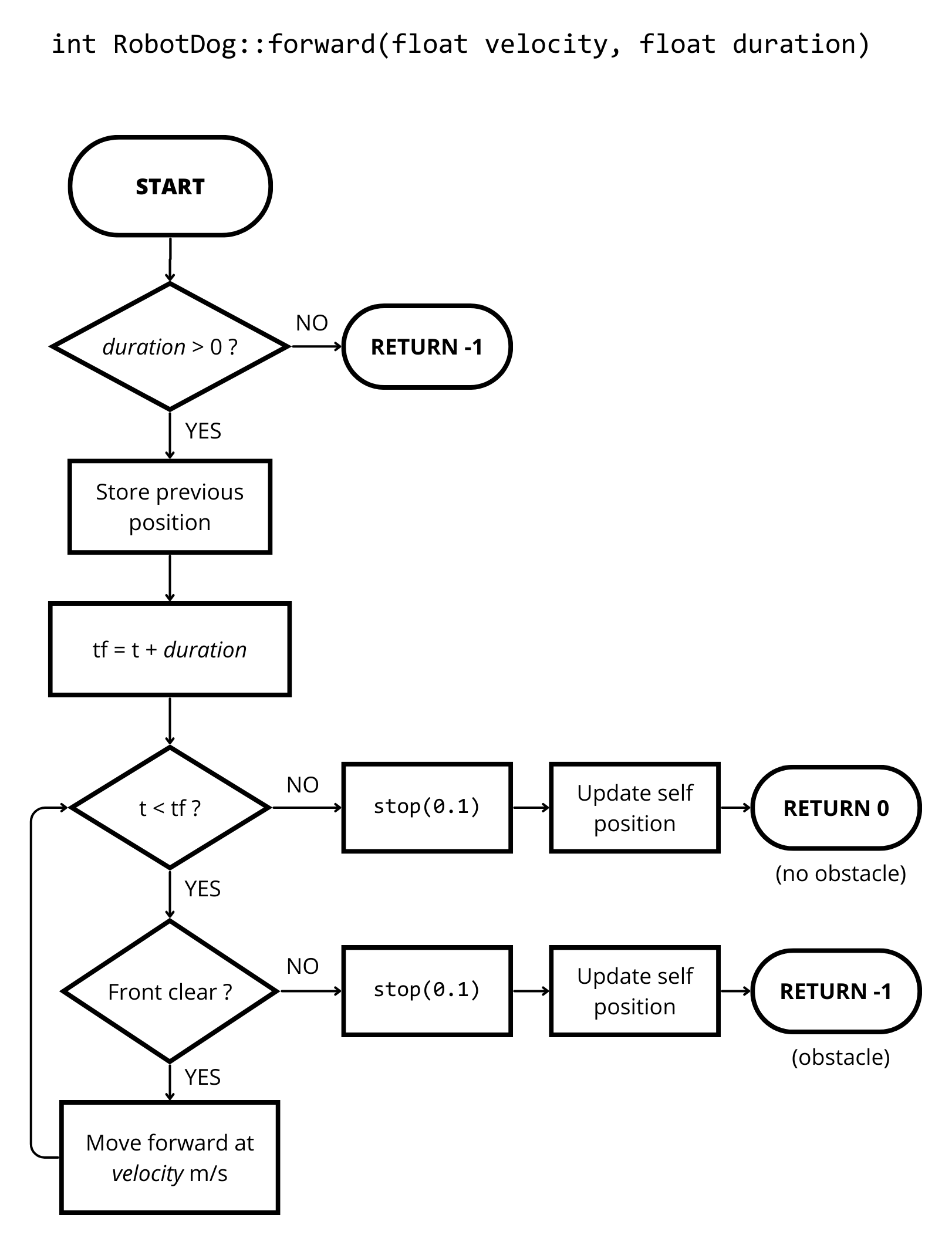} }}%
        \quad
        \subfloat[\centering Function $stop$.]{{\includegraphics[width=0.45\columnwidth]{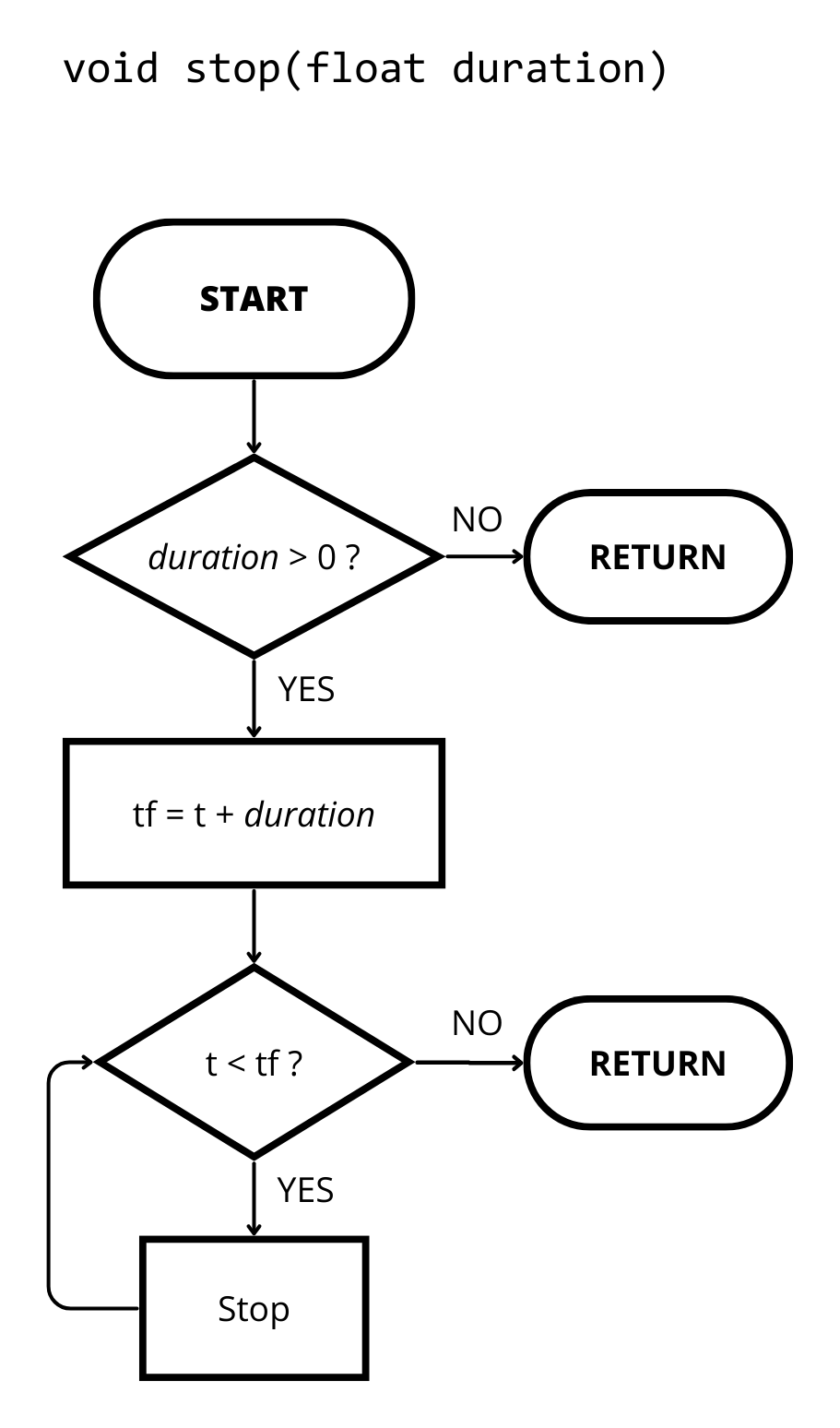}}}%
        \qquad
        \subfloat[\centering Function $rotate$.]{{\includegraphics[width=0.45\columnwidth]{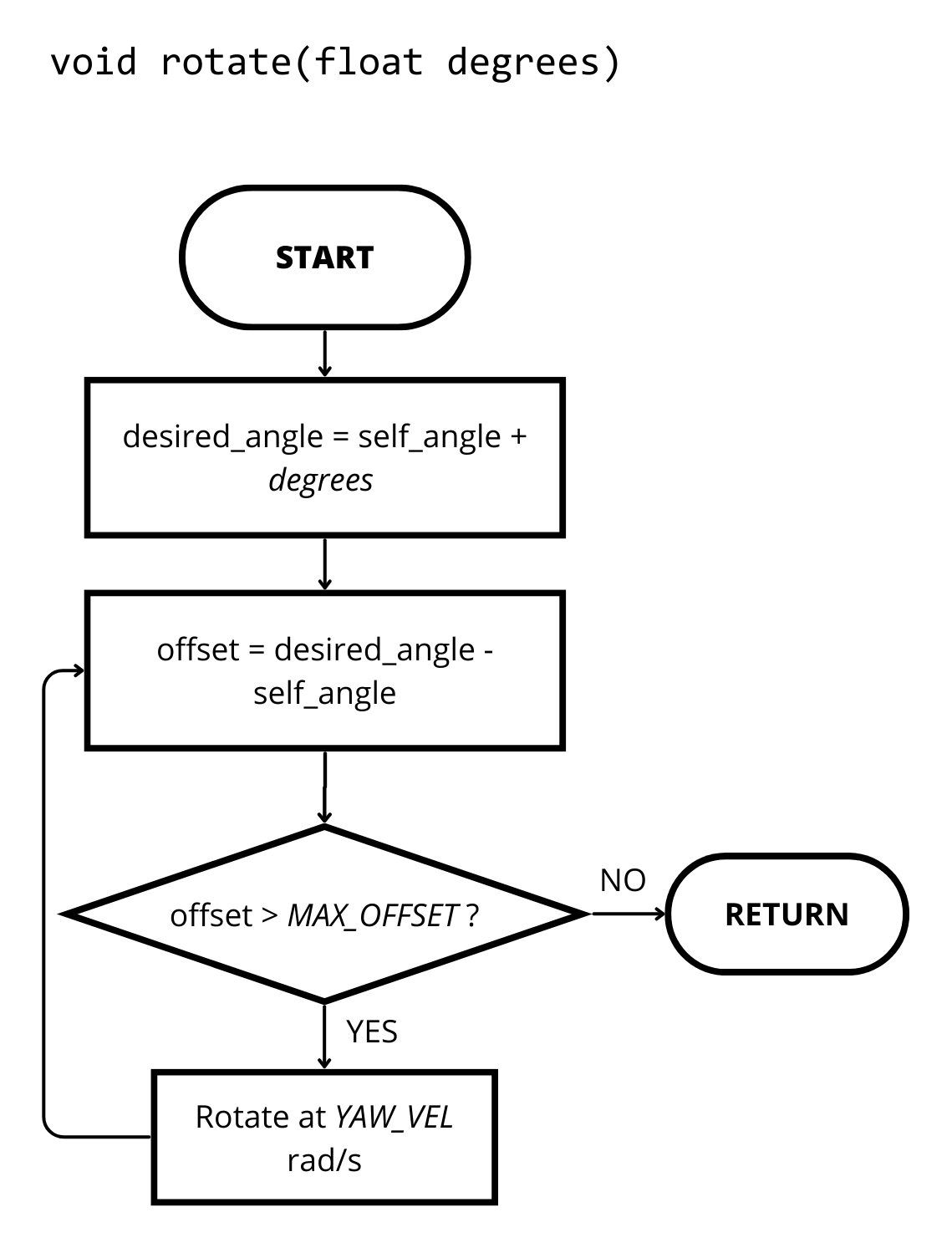} }}%
    \caption{Control Module's functions that are used to control the movement of the MRP.}%
    \label{fig:flux_for_rot}%
\end{figure}

The Control Module was implemented building upon \cite{ros2real} and \cite{math}, which are implementations provided by Unitree developers, based on the Robot Operating System (ROS). The Control Module's operation consists of different functions, including $forward$, $stop$, and $rotate$. The first function allows calculating the movement duration, based on the current time and the predefined time interval provided as a parameter, as well as controlling the MRP movement during that interval. If an obstacle is detected along the movement trajectory, the $stop$ function is called and a value representing an obstacle detection is returned. The $rotate$ function allows calculating the MRP's orientation and rotating the MRP accordingly, in order to enable reaching the targeted position. The logic behind the $forward$, $stop$, and $rotate$ functions that make part of the Control Module is presented in \cref{fig:flux_for_rot}.\looseness=-1 

\begin{figure}[!ht]
  \begin{center}
    \includegraphics[width=0.7\columnwidth]{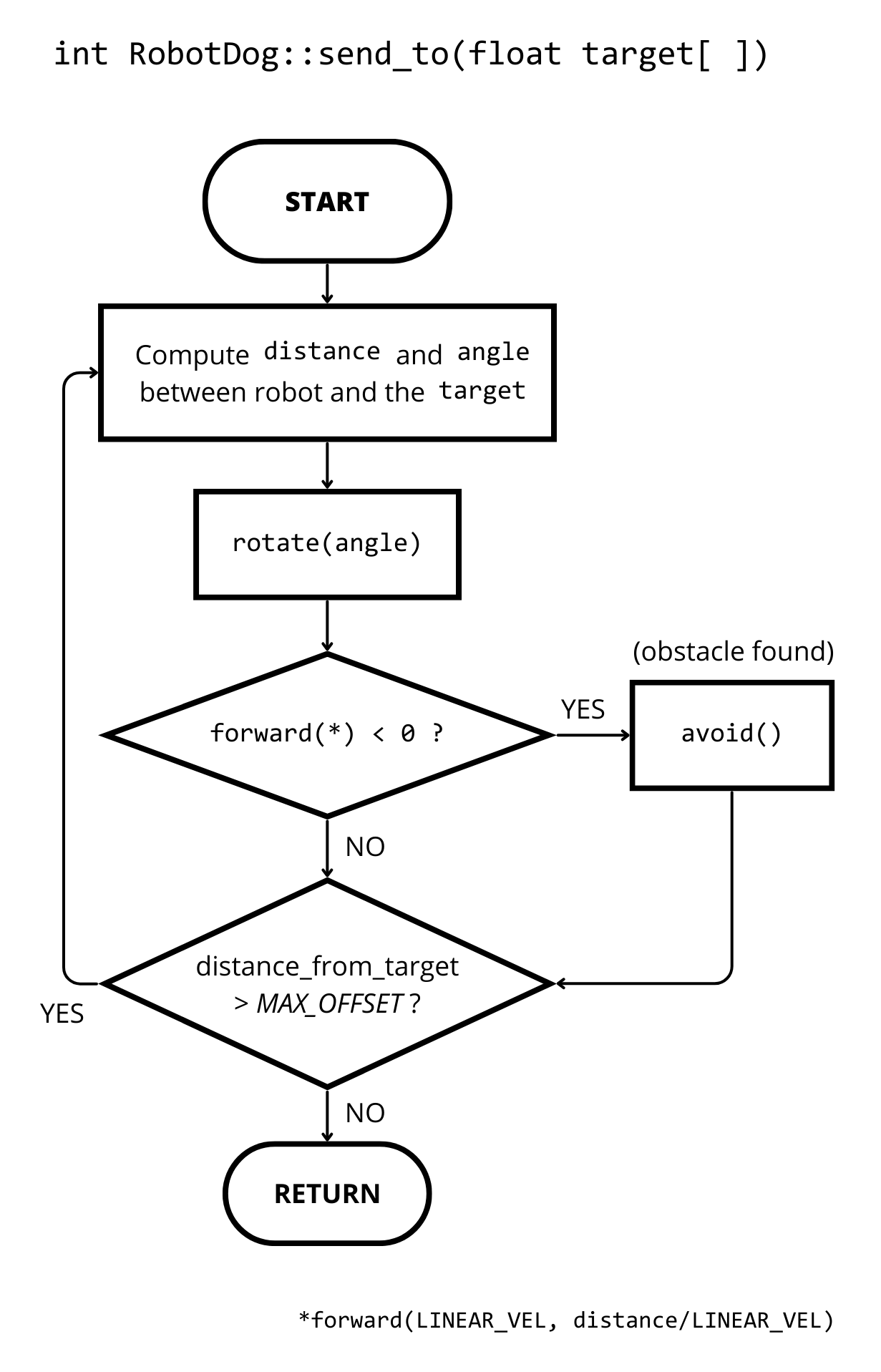}
    \caption{Control Module's $send\_to$ function, which employs the $forward$, $rotate$, and $avoid$ functions.}
    \label{fig:send_to}
  \end{center}
\end{figure}

\begin{figure}[!ht]
  \begin{center}
    \includegraphics[width=0.7\columnwidth]{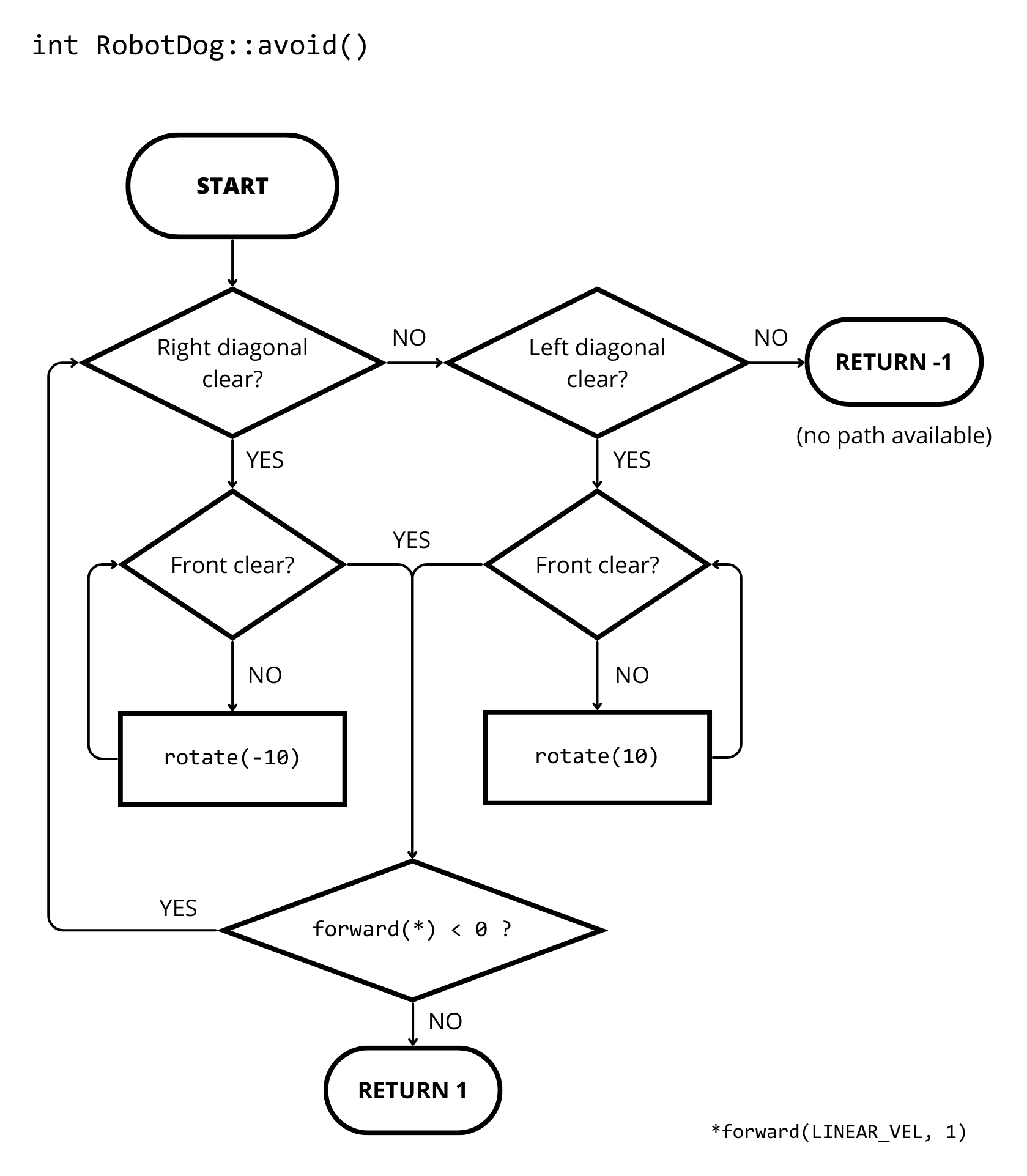}
    \caption{Control Module's $avoid$ function.}
    \label{fig:flux_avoid}
  \end{center}
\end{figure}

In order to control the movement of the MRP taking advantage of the $rotate$ and $forward$ functions, the $send\_to$ function, depicted in \cref{fig:send_to}, was implemented. It allows computing the orientation angle and distance to the targeted position. When the $forward$ function returns a negative value, which occurs when an obstacle is detected, the $avoid$ function is called. When the targeted position is reached by the MRP, if there is a significant deviation in practice, the $send\_to$ function is called recursively; otherwise, the process is concluded. The $avoid$ function, shown in \cref{fig:flux_avoid}, is in charge of deviating the MRP from the obstacles that are detected during its movement. When called, the $avoid$ function triggers continuously the $rotate$ function with a default angle (in the right-hand side first). When no obstacles are detected in the MRP's front, the $forward$ function is called with a default speed and duration time. If the movement is successful, the function returns. If not, it means that another obstacle was found, and the $avoid$ function is called recursively.\looseness=-1 

In order to exchange control information between the Control Module and the MRP an UDP-based connection is employed. The Control Module receives information from the MRP about its position, Inertial Measurement Unit (IMU), velocity, and distance to potential obstacles in left, right and front sides. Moreover, the Control Module receives input data from the Vision Module, which allows us to estimate the relative positions of obstacles. By cross-referencing this information with the distance information provided by the ultrasonic sensors of the MRP, it is possible to characterize the surrounding environment and infer if an obstacle is obstructing the MRP trajectory and the wireless signal propagation.\looseness=-1 

\section{Evaluation\label{sec:SystemValidation}}

In order to validate the proposed solution, the Control Module was tested using Turtlesim \cite{turtlesim}. Turtlesim is a 2D simulator, compatible with ROS, that allows building a virtual environment for controlling an MRP using ROS-based movement commands.\looseness=-1 

\begin{figure*}[!ht]
\centering
        \includegraphics[width=0.7\textwidth]{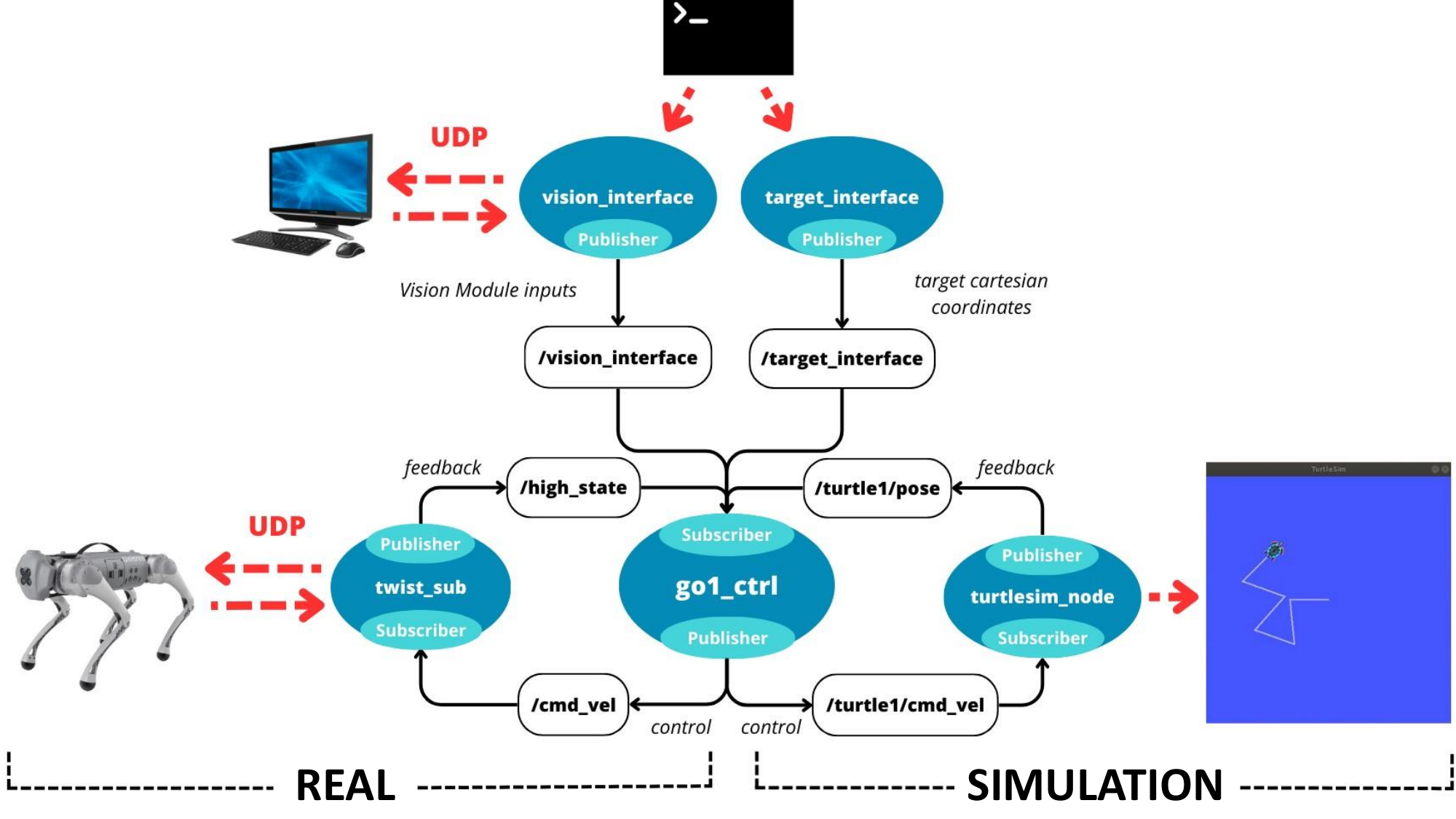}
    \caption{Architecture of the ROS-based system designed to validate the Vision Module and the Control Module.}
    \label{fig:ros_diagram}
\end{figure*}

The architecture of the ROS-based system designed to validate the Vision Module and the Control Module is depicted in \cref{fig:ros_diagram}. It includes a main node, $go1\_ctrl$, where the Control Module is running, while communications with the remaining system's nodes are established. Control instructions are simultaneously sent by the main node and received by the nodes $twist\_sub$ and $turtlesim\_node$. The node $twist\_sub$ is responsible for establishing an UDP-based connection with the MRP and formatting ROS-based messages into instructions accepted by the MRP. The $turtlesim\_node$ is responsible for exchanging information with Turtlesim. The system's user can choose which of these nodes are used to extract the position feedback information. A terminal-based interface was also developed. This interface allows the user to manually introduce information provided by the Vision Module, in case the Control Module is used independently. As such, the $vision\_interface$ node can be created by the terminal-based interface or when launching the UDP-based server that allows receiving real-time data provided by the Vision Module. The $target\_interface$ node is used to send the target MRP's position coordinates to the Control Module. The Control Module uses this information to move the MRP to the desired position.\looseness=-1 

\subsection{Simulation Results}
\begin{figure}[!ht]
    \centering
        \subfloat[First trajectory without obstacles.\label{fig:testA}]{{\includegraphics[width=0.44\columnwidth]{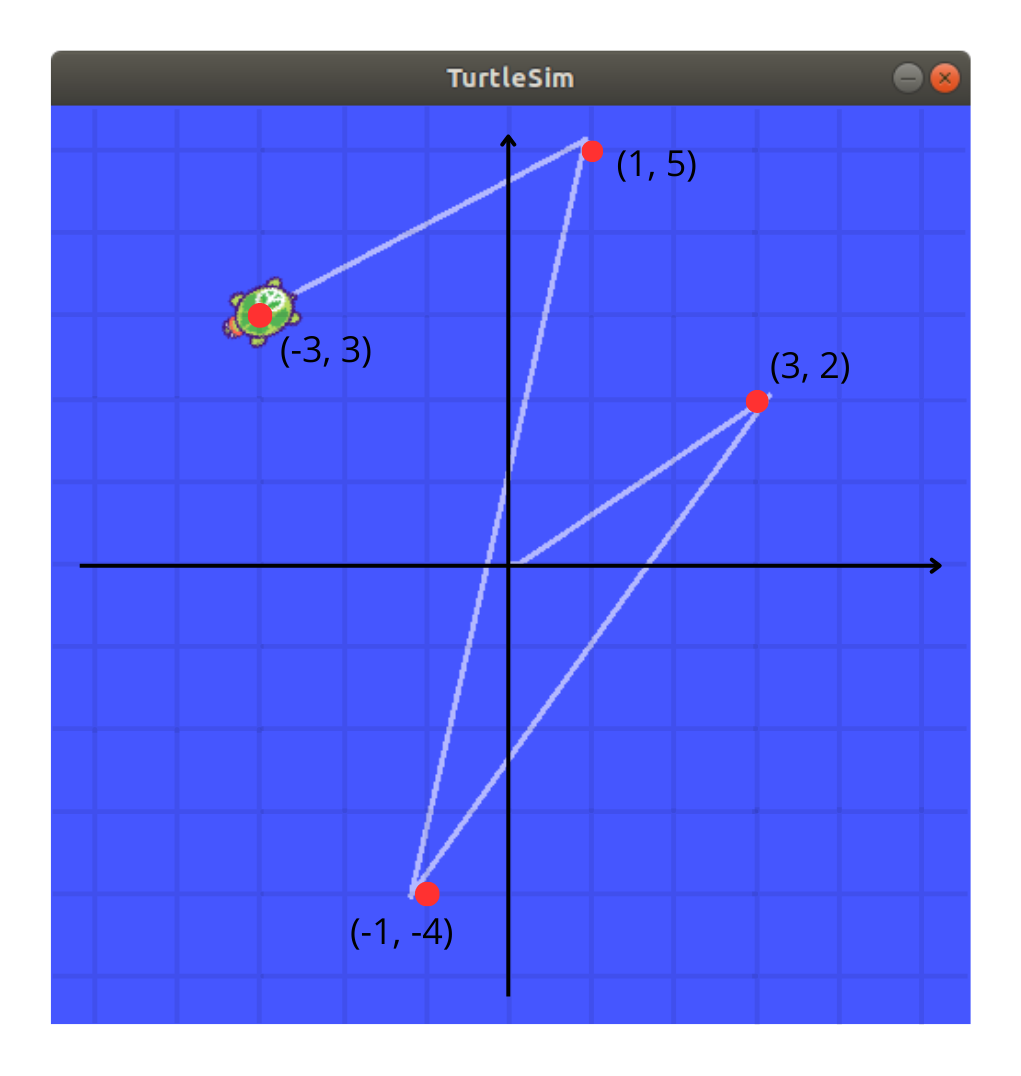} }}%
        \hfill
        \subfloat[Second trajectory without obstacles.\label{fig:testB}]{{\includegraphics[width=0.44\columnwidth]{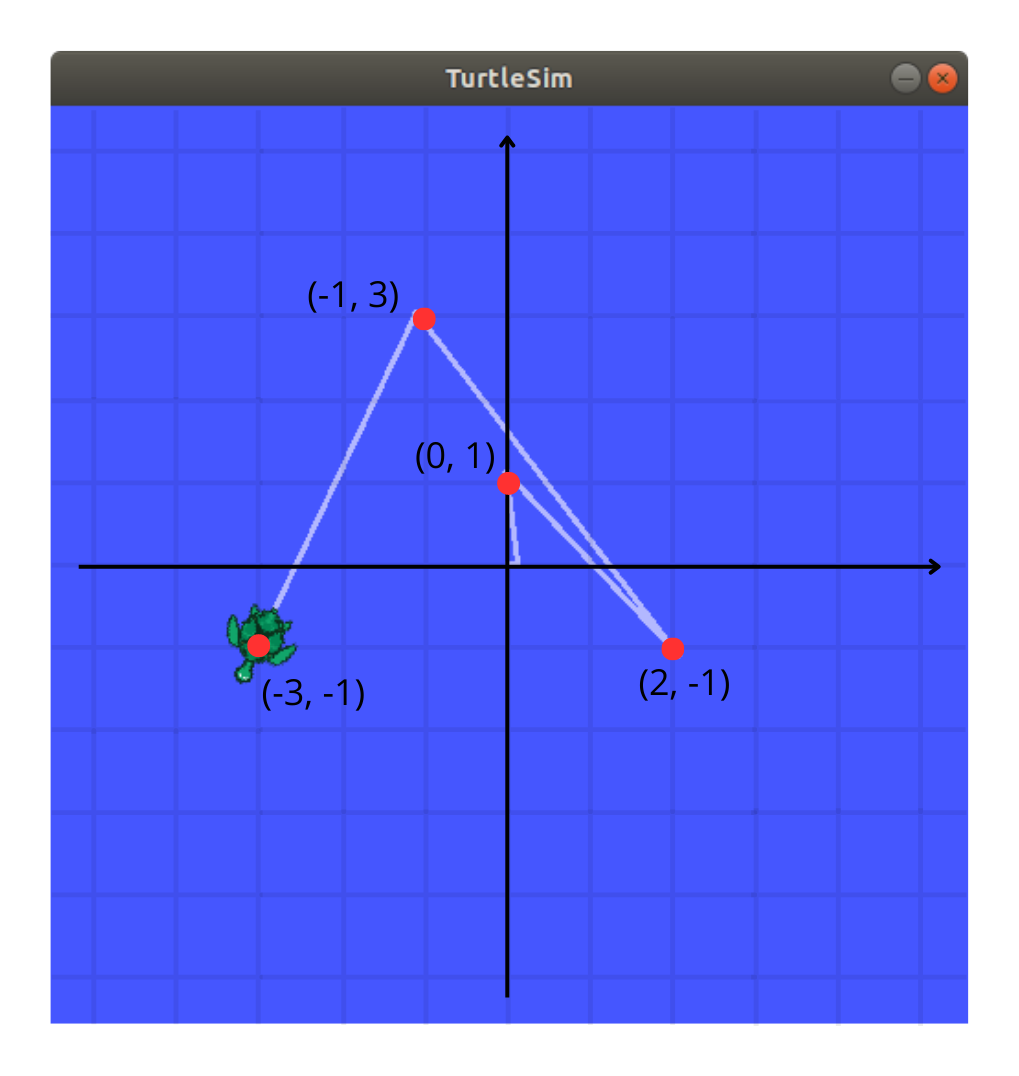} }}%
        \hfill
        \subfloat[Trajectory with an obstacle (the red square), in order to validate the obstacle avoidance mechanism of the Control Module.\label{fig:testC}]{{\includegraphics[width=0.44\columnwidth]{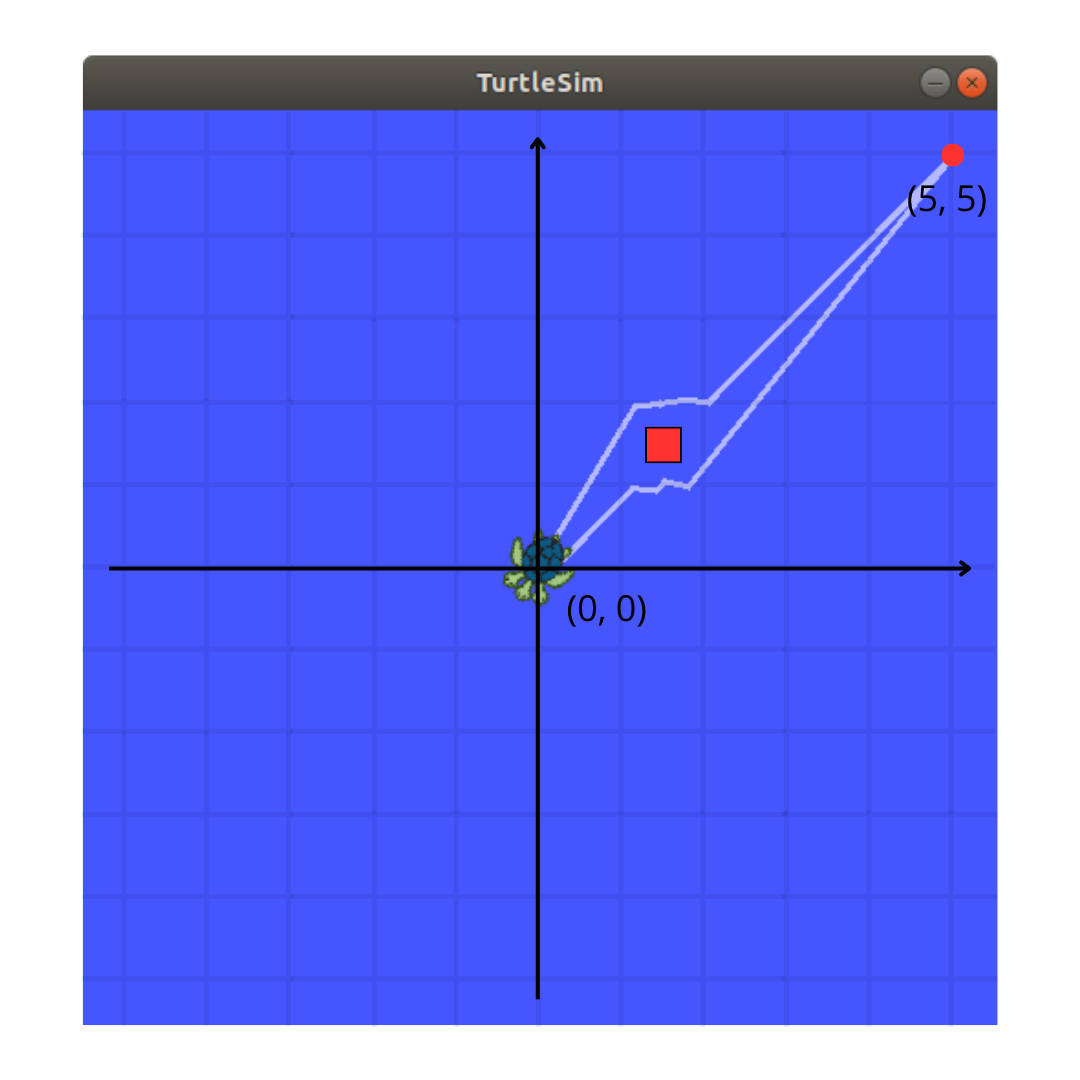} }}%
    \caption{Simulation results.}
    \label{fig:turtlesim}
\end{figure}

The simulation results regarding the validation of the Control Module using Turtlesim are presented in \cref{fig:turtlesim}. The results presented in \cref{fig:testA} and \cref{fig:testB} consider different trajectories for the MRP, defined by means of Cartesian coordinates, starting the movement at the origin represented by the turtle symbol; these tests aimed at validating the successful control of the MRP towards a targeted position. \cref{fig:testC} shows the validation results for the obstacle avoidance mechanism of the Control Module, considering a back-and-forth movement between position $(0, 0)$ and position $(5, 5)$. In the simulation tests used to validate the Control Module, the terminal-based interface was used to provide information about obstacles in the MRP's trajectory. In a real-world scenario, this information is provided by the Vision Module.\looseness=-1 

\subsection{Experimental Results}
The experimental tests considered the joint validation of the Control Module and the Vision Module, including the classification of objects detected in video frames, their position, and the transmission of the resulting information to the Control Module. The experimental scenario considered an obstacle blocking the LoS to the device seeking wireless connectivity. The obstacle was a plant vase and the device was abstracted by a person.\looseness=-1 

\begin{figure}[!ht]
	\begin{center}
		\includegraphics[width=0.7\columnwidth]{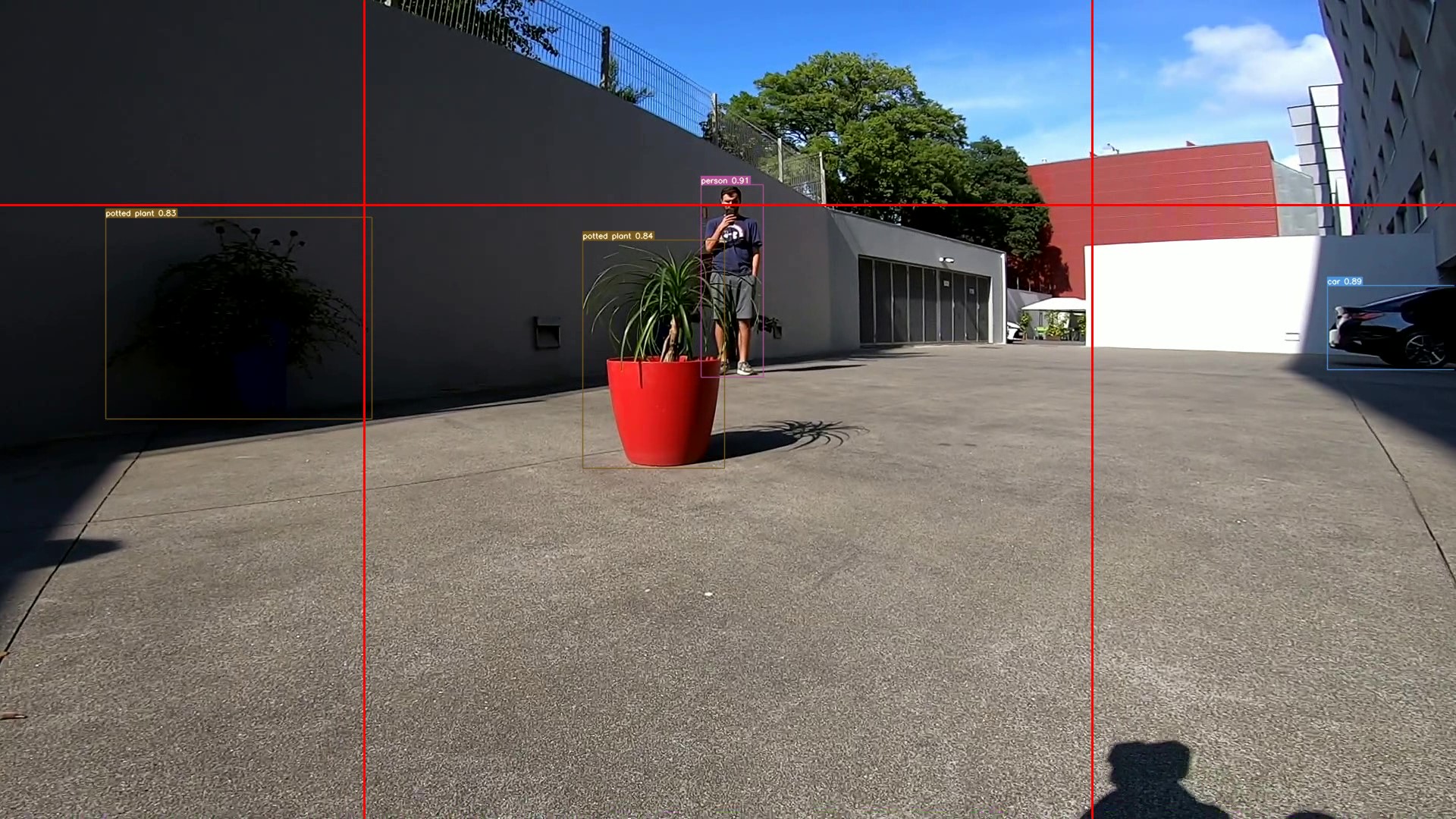}
		\caption{MRP's view before starting the movement.}
		\label{fig:video2_1}
	\end{center} 
\end{figure}

\begin{figure}[!ht]
	\begin{center}
		\includegraphics[width=0.7\columnwidth]{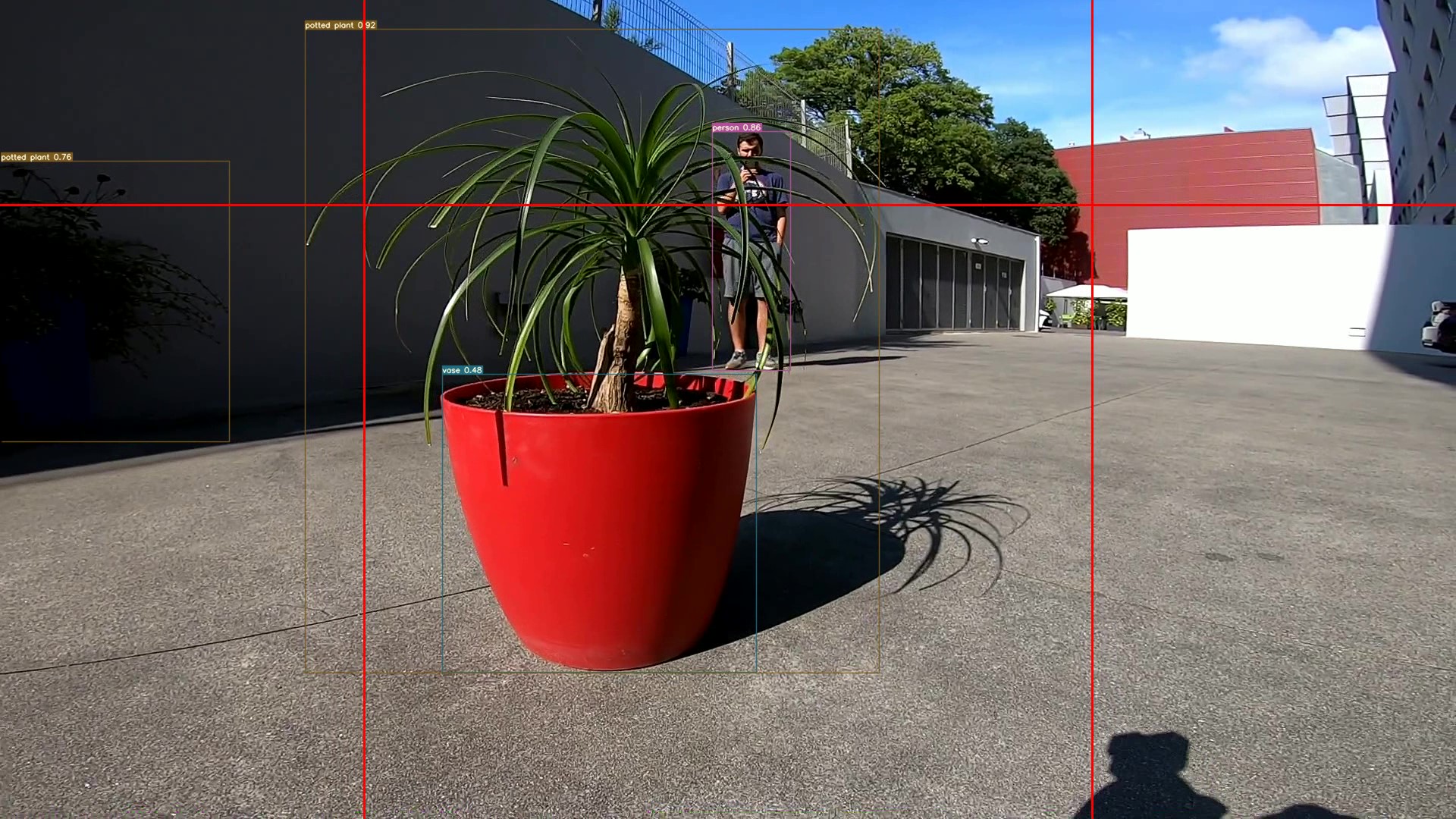}
		\caption{MRP's view at the moment when the plant vase is detected as an obstacle to avoid.}
		\label{fig:video2_2}
	\end{center} 
\end{figure}

\begin{figure}[!ht]
	\begin{center}
		\includegraphics[width=0.7\columnwidth]{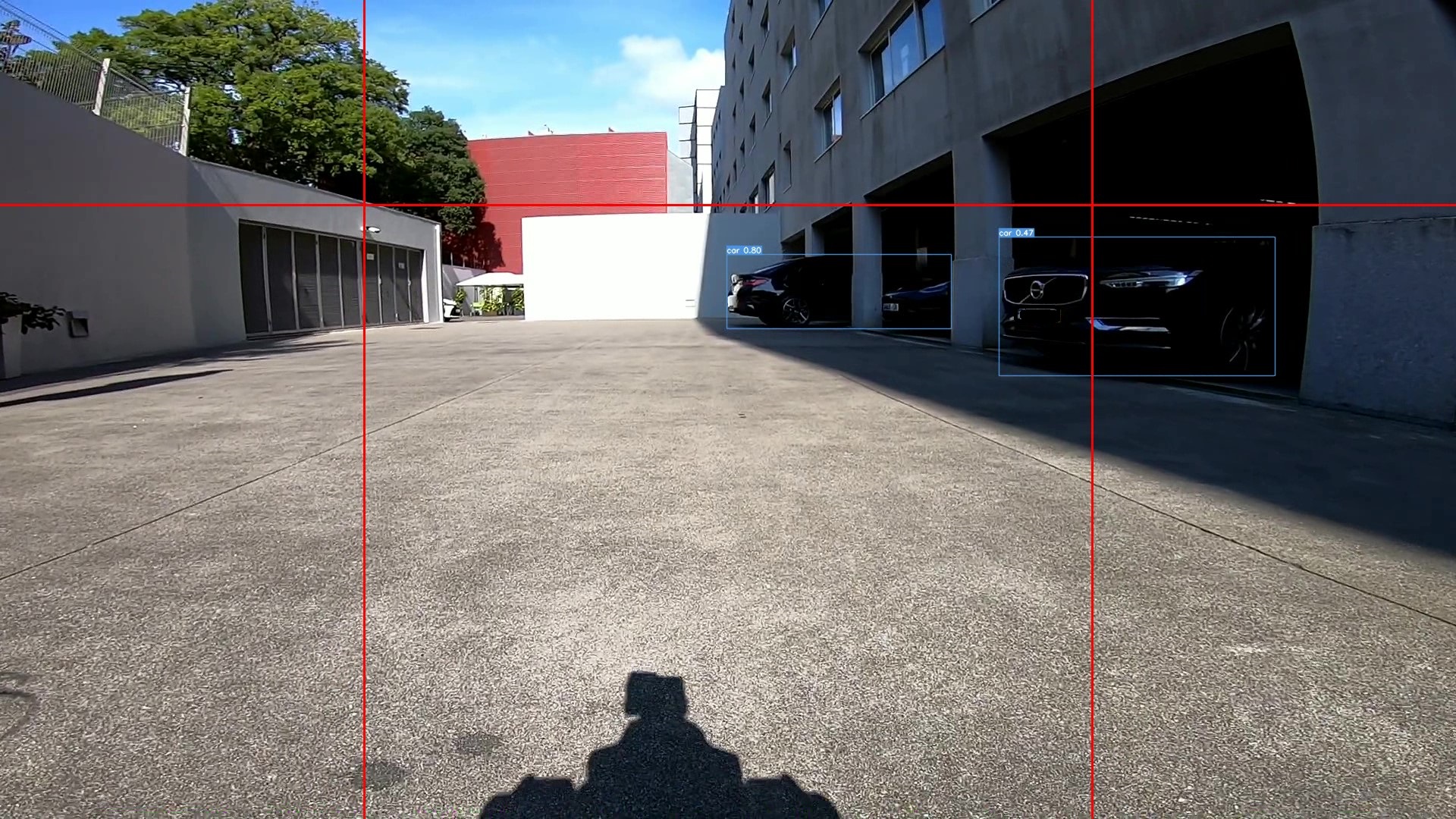}
		\caption{Moment when any obstacle is no longer within the MRP's field of vision.}
		\label{fig:video2_3}
	\end{center} 
\end{figure}

\begin{figure}[!ht]
	\begin{center}
		\includegraphics[width=0.7\columnwidth]{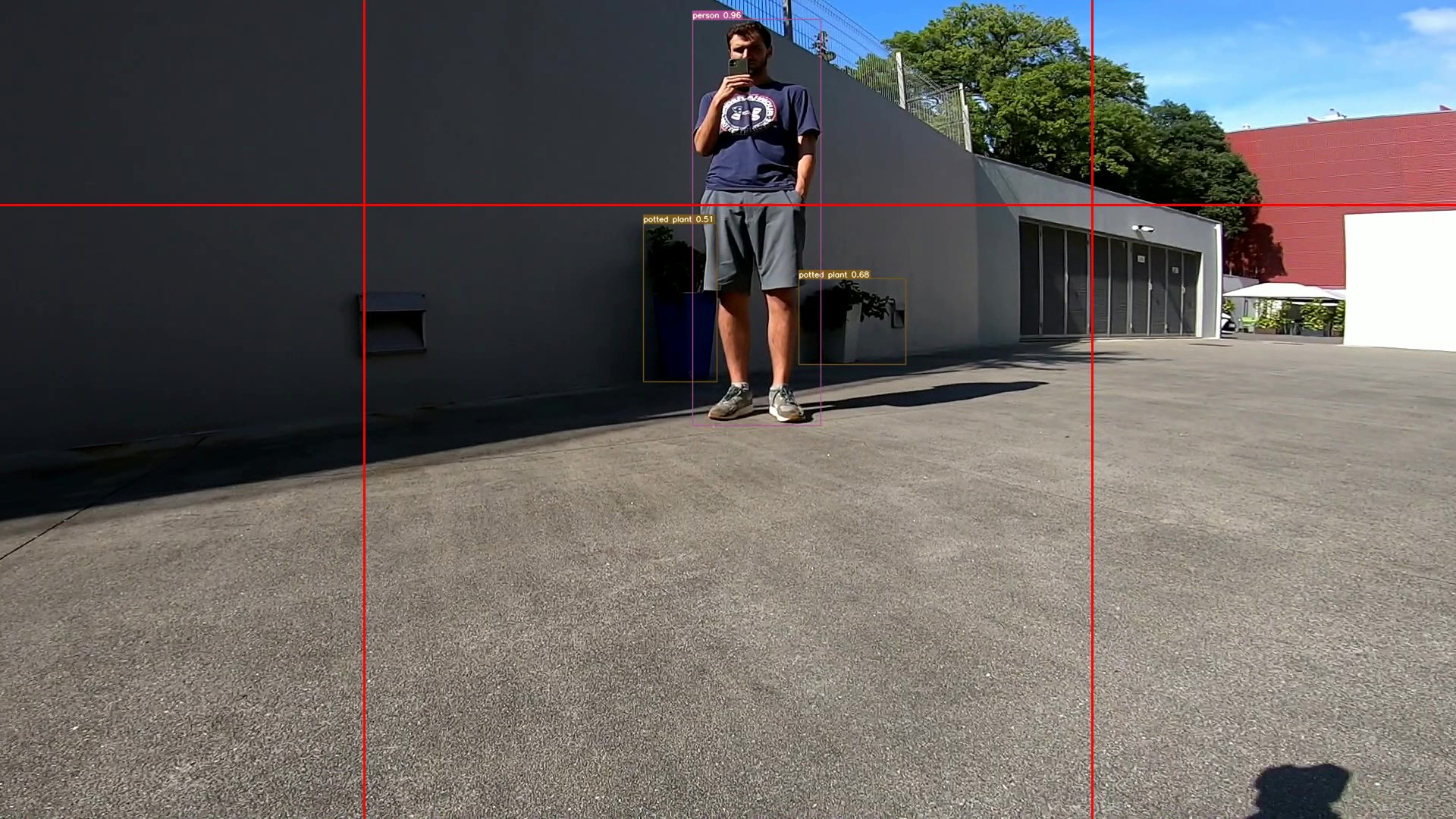}
		\caption{MRP's view after the avoidance of the obstacle.}
		\label{fig:video2_4}
	\end{center} 
\end{figure}

In the experimental testing carried out, the Vision Module was able to detect the obstacle through the video. Moreover, upon receiving the distance information provided by the MRP's ultrasonic sensors, it allowed determining whether it was necessary to deviate from the obstacle or not, by inferring if the obstacle was close enough of the MRP. The resulting information was sent to the Control Module, which controlled the movement of the MRP to overcome the obstacle. The complete footage of the experimental testing carried out can be viewed in \cite{video2}. \cref{fig:video2_1} shows the position of the MRP before starting the movement. \cref{fig:video2_2} depicts the moment when the obstacle was detected, leading the Control Module to redefine the trajectory path to avoid the obstacle. In Figure \ref{fig:video2_3}, the MRP has deviated enough for the obstacle to be out of its field of view. Therefore, information was sent by the Control Module to the MRP so that the movement was resumed. Finally, in \cref{fig:video2_4}, the MRP has resumed the initial trajectory, with the device centered in its field of view.\looseness=-1 
 
\section{Discussion \label{sec:Discussion}}
The experimental evaluation carried out allowed validating that it is possible to autonomously detect and avoid obstacles that hinder the establishment of LoS wireless links between a mobile communications cell and wireless devices, using the information obtained through video captured by the MRP. By ensuring LoS wireless links, the QoS offered can be maximized while avoiding communications disruptions in on-demand wireless networks.\looseness=-1 

Despite the successful validation achieved, the proposed solution has some limitations, which may be addressed in future works. First, real-time operation is constrained by the high computational requirements of the Vision Module, especially due to the use of the YOLO algorithm for object detection. The integration of a GPU with increased capabilities will potentially improve the efficiency of real-time object detection and classification. Moreover, the employed approach for object detection and classification is relatively limited, which may result in less accurate results under certain conditions, especially when considering small size and moving obstacles. This limitation is caused by the finite dataset used for training the model employed by the YOLO algorithm, which may lead to inefficiency in detecting some objects. Future works may consider optimizing the proposed solution to use less powerful graphics cards. Moreover, training a more suitable model for representative wireless networking scenarios may improve object detection capabilities, while enhancing the overall performance of the system in practice.\looseness=-1 

Despite the existing limitations, the evaluation carried out in simulation and in the real-world has demonstrated the ability to avoid obstacles while enabling the autonomous control of the MRP, taking into account video information to characterize the surrounding environment.\looseness=-1 

\section{Conclusions\label{sec:Conclusions}}
The emergence of the 6G paradigm introduces the need of flexible wireless network infrastructures able to support a massive number of interconnected devices. For that purpose, the integration of MRPs into 6G networks is a promising approach to enable the creation of on-demand mobile communications cells. However, the challenge lies in positioning the MRPs to improve the wireless connectivity offered.\looseness=-1 

This paper proposed a novel solution for the obstacle-aware autonomous positioning of MRPs to provide LoS wireless connectivity to devices and users. It consists of a framework including a Vision Module and a Control Module. The Vision Module uses video data to identify the location of obstacles and wireless devices, while the Control Module uses this information to autonomously position the MRP. The proposed solution was successfully validated in simulation and through experimental testing.\looseness=-1 

As future work it is worth considering improving the accuracy of object detection and classification, taking advantage of models optimized for typical wireless networking scenarios, and assessing the scalability of the proposed solution, including the support for multiple MRPs.\looseness=-1 

\section*{Acknowledgments}
This work is financed by National Funds through the Portuguese funding agency, FCT -- Fundação para a Ciência e a Tecnologia, within project UIDB/50014/2020.

\printbibliography

\end{document}